\documentstyle[epsfig]{elsart}
\newcommand{\gsim}{\mathrel{\rlap{\lower4pt\hbox{\hskip1pt$\sim$}}
\raise1pt\hbox{$>$}}}
\newcommand{\lsim}{\mathrel{\rlap{\lower4pt\hbox{\hskip1pt$\sim$}}
\raise1pt\hbox{$<$}}}
\newcommand{\sfrac}[2]{\mbox{\footnotesize $\frac{#1}{#2}$}}
\begin{document}
\begin{frontmatter}
\hspace*{\fill}{Preprint Numbers: \parbox[t]{100mm}{ANL-PHY-8678-TH-97
        \hspace*{\fill} nucl-th/9704039}}

\title{Semileptonic decays of heavy mesons}
\author[dubna]{M. A. Ivanov,}
\author[dubna1]{Yu. L. Kalinovsky,}
\author[anl]{P. Maris,}
\author[anl]{and C. D. Roberts}
\address[dubna]{Bogoliubov Laboratory of Theoretical Physics, \\
Joint Institute for Nuclear Research, 141980 Dubna, Russia}
\address[dubna1]{Laboratory of Computing Techniques and Automation, \\
Joint Institute for Nuclear Research, 141980 Dubna, Russia}
\address[anl]{Physics Division, Bldg. 203, Argonne National Laboratory,\\
Argonne IL 60439-4843, USA}
\begin{abstract}
The semileptonic and leptonic decays of heavy mesons are studied as a
phenomenological application and exploration of a heavy-quark limit of
Dyson-Schwinger equations.  The single form factor, $\xi(w)$, which
characterises the semileptonic decay in this limit, is calculated and
compares well with recent experimental extractions.  We obtain a lower bound
of $1/3$ on the slope-parameter $\rho^2 \equiv -\xi^\prime(1)$, which, in
calculations with realistic input, is exceeded by a great deal: agreement
with experimental data requiring $\rho^2 \sim 1.2\,$-$\,1.6$.  The flavour
and momentum dependence of the light-quark propagators has observable
consequences.
\end{abstract}
\end{frontmatter}

{\bf 1. Introduction}.\\ Semileptonic decays of pseudoscalar mesons provide a
means of measuring elements of the Cabibbo-Kobayashi-Maskawa (CKM) matrix;
fundamental parameters of the standard model.  For example: the $K_{e 3}$
decay, $K\to \pi e \nu_e$, can be used to determine $|V_{us}|$; $D_{e 3}$ to
determine $|V_{cs}|$; and semileptonic decays of $B$ mesons can provide
information about $|V_{cb}|$ and $|V_{ub}|$.  With a single hadron in both
the initial and final state, these processes are an ideal tool for studying
the influence of nonperturbative, strong interaction dynamics on weak
interaction processes.  An accurate determination of the CKM matrix elements
relies on developing a good understanding of these nonperturbative effects.

The simplest of these processes to study are those with a pseudoscalar meson
in both the initial and final state.  In this case the interaction is only
sensitive to the vector part of the electroweak $(V-A)$ interaction and, of
the two form factors, denoted $f_\pm(t)$, $f_+(t)$ is dominant for $e$ and
$\mu$ final states, for which the leptonic current is approximately
conserved.  In theoretical studies of these decays of light pseudoscalar
mesons, $\pi$ and $K$, nonperturbative dressing of the quark-W-boson vertex
is important, with nonanalytic contributions associated with, for example,
$K$-$\pi$ loops, being significant for $t\gsim 0$~\cite{KMR97}.  For heavy
mesons, e.g. $D$ and $B$, such nonanalytic contributions are unimportant in
the physical region: $0\lsim t \leq (m_B-m_D)^2$.  However, this does not
automatically mean that one can use the bare vertex, $\gamma_\mu$, because
these are not the only dressing effects induced by the strong interaction.

\begin{figure}[t]
\centering{\
\epsfig{figure=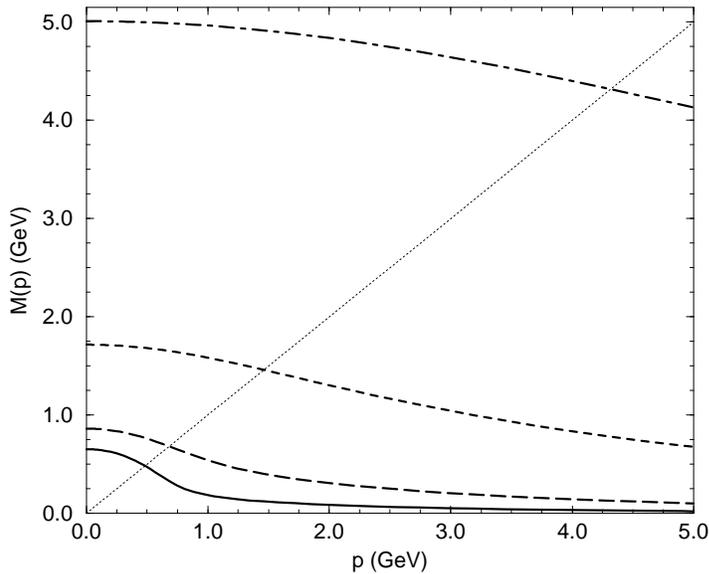,height=9.0cm}}
\caption{The behaviour of $M_f(p^2)$: $u/d$-quark - solid line; $s$-quark -
long-dashed line; $c$-quark - dashed line; and $b$-quark - dot-dash line,
obtained using the model of Ref.~\protect\cite{FR96}.  See also
Refs.~\protect\cite{WKR91,MJ92}.  The diagonal dotted line is $p=M(p)$.
\label{figqp}}
\end{figure}

The propagator for a dressed quark of flavour $f$ can be written\footnote{We
employ a Euclidean space formulation with
$\{\gamma_\mu,\gamma_\nu\}=2\delta_{\mu\nu}$, $\gamma_\mu^\dagger =
\gamma_\mu$ and $a\cdot b=\sum_{i=1}^4 a_i b_i$.  A timelike vector, $Q_\mu$,
has $Q^2<0$.}
\begin{eqnarray}
S_f(p) & = & -i\gamma\cdot p \sigma_V^f(p^2) + \sigma_S^f(p^2)
=
\frac{Z_f(p^2)}{i\gamma\cdot p + M_f(p^2)}\,.
\end{eqnarray}
The characteristic behaviour of $S_f(p)$ in QCD can be estimated by studying
the quark Dyson-Schwinger equation (DSE)~\cite{DSErev} and is illustrated in
Fig.~\ref{figqp}.  As typical of these calculations, the different quark
flavours are specified by the value of the current-quark mass at the
renormalisation point, $\mu\simeq 10\,$GeV in this heuristic illustration
using the simple model of Ref.~\cite{FR96}, with (in MeV): $m_u(\mu)\approx$
$1$; $m_s(\mu)\approx$ $25$; $m_c(\mu)\approx$ $300$; and $m_b(\mu)\approx$
$3000$.  The intersection of the dotted, diagonal line with a given curve
marks the solution of $p=M_f(p)$ and defines the Euclidean constituent quark
mass, $M_f^E$.\footnote{Confinement entails that there is no real solution of
$p^2+M_f(p^2)=0$; i.e., no ``pole-mass'' for the quarks.}  The ratio
$M_f^E/m_f$ is a single, indicative and quantitative measure of the
nonperturbative effects of gluon-dressing on the quark propagator.

Obvious in this figure is a signal difference between light- and
heavy-quarks: for light-quarks, $M_f^E/m_f \gsim 20$, while, for
heavy-quarks, $M_f^E/m_f \lsim 5$.  This is a manifestation of the fact that
the heavy-quark mass functions are well represented by a constant whereas
those of light-quarks are not; the accuracy of this approximation improving
with increasing current-quark mass.  The same is true of $Z_f(p^2)$; i.e.,
$Z_{f_Q}(p^2)\approx 1$.  This feature provides the basis for a
simplification of our study of semileptonic decays of heavy mesons: {\it in
the kinematic region explored by the decays} one can approximate the dressed
heavy-quark propagator as
\begin{equation}
\label{hqa}
S_{f=c,b} \approx \frac{1}{i \gamma\cdot p + \hat M_f}\,,
\end{equation}
where $\hat M_f \approx M_f^E$; i.e., in the first instance, one can ignore
the momentum dependence of gluon dressing of the heavy-quark propagator.  In
the DSE framework, ``heavy-quark symmetry'' \cite{IW90} finds its foundation
in this.

Returning to the dressed quark-W-boson vertex, which describes the coupling
of dressed-quarks to the W-boson, the vector piece satisfies the identity
\begin{equation}
\label{wti}
Q_\mu i V_\mu^{f_1 f_2} (p;Q) = 
S^{-1}_{f_1}(p_+) - S^{-1}_{f_2}(p_-) 
- (m_{f_1} - m_{f_2}) \Gamma_I^{f_1 f_2} (p;Q)\,,
\end{equation}
where $p_+ = p + \eta\, Q$, $p_- = p - \hat\eta\, Q$, $\hat\eta = (1-\eta)$,
and $Q$ is the total momentum.  Here $\eta$ is the momentum partitioning
parameter that arises because of the arbitrariness in the definition of the
relative momentum in a covariant formalism, and $\Gamma_I^{f_1 f_2} (p;Q)$ is
the flavour-dependent scalar vertex.  (In the absence of interactions
$\Gamma_I^{f_1 f_2} (p;Q) = I_D$.)  In studying the semileptonic decays of
light mesons, the dressing of the vertex implied by this Ward-Takahashi
identity was itself important, even neglecting the nonanalytic contributions
mentioned in the introduction~\cite{KMR97}.  However, in the case of
heavy-quarks, the ability to neglect the momentum dependence of gluon
dressing entails Eq.~(\ref{hqa}) and $(m_{f_1} - m_{f_2})\Gamma_I^{f_1 f_2}
(p;Q) \approx (\hat M_{f_1} - \hat M_{f_2}) I_D$; and hence Eq.~(\ref{wti})
is satisfied approximately by the bare vertex.  This justifies the
approximation
\begin{equation}
\label{hqvtx}
V_\mu^{f_1 f_2} (p;Q) = \gamma_\mu\,,
\end{equation}
amplifying the heavy-quark simplification in the study of these decays.

{\bf 2. Semileptonic and leptonic decays.}\\ Herein we study the impulse
approximation to the semileptonic $B^0\to D^- \ell^+ \nu$ decay amplitude,
defined by
\begin{eqnarray}
\label{semilep}
\lefteqn{\langle D^-(K) | \bar b \gamma_\mu c | B^0(P) \rangle
\equiv f_+(t) (K+P)_\mu - f_-(t) (K-P)_\mu}\\
&& = \nonumber
N_c \int\frac{d^4 \ell}{(2\pi)^4}
{\rm tr}_D\left[
\bar\Gamma_{D^-}\left(\ell + \hat\eta P; - K\right)
S_d(\ell + \hat\eta (P+K) )\Gamma_{B^0}\left(\ell + \hat\eta K ;P\right)
\times
\right.\\
&&\nonumber\left.
S_b(\ell - \eta P + \hat\eta K)
iV_\mu^{bc}(\ell+ (1-2 \eta) K;K-P)
S_c(\ell+ \hat\eta P - \eta K )\right]\,,
\end{eqnarray}
where $t= - (P-K)^2$, $\bar \Gamma_{B^0,D^-}(k;P)^{\rm T} =
C^\dagger\Gamma_{B^0,D^-}(-k;P)C$, $C=\gamma_2\gamma_4$ is the charge
conjugation matrix, and $\Gamma_{B^0,D^-}$ are the meson Bethe-Salpeter
amplitudes,\footnote{In solving a Bethe-Salpeter equation to obtain the bound
state mass and $\Gamma$, the $\eta$-dependence of $\Gamma$ ensures that
physical observables, such as the mass, are independent of $\eta$.}
normalised via
\begin{eqnarray}
\nonumber
&& N_c \int\frac{d^4 k}{(2\pi)^4}
{\rm tr}_D\left[
\bar \Gamma_{B^0,D^-}(k;-P)\partial_\mu^P S_{d}(k+\hat\eta P)
        \Gamma_{B^0,D^-}(k;P) S_{b/c}(k-\eta P)\right.\\
&&+ \left.\bar \Gamma_{B^0,D^-}(k;-P)S_{d}(k+\hat\eta P)
        \Gamma_{B^0,D^-}(k;P) \partial_\mu^P S_{b/c}(k-\eta P)\right]
= 2 P_\mu\,,
\label{bsanorm}
\end{eqnarray}
which is the consistent, canonical normalisation in impulse
approximation~\cite{CDRpion}.

Simultaneously we study the leptonic decay of a heavy, pseudoscalar meson,
$M$, which is described by one, dimensioned coupling, $f_M$:
\begin{eqnarray}
\label{lep}
\lefteqn{\langle 0 | \bar f_2 \gamma_\mu \gamma_5 f_1| \Phi_M(P)\rangle 
\equiv}\\
\nonumber
&& f_M P_\mu = N_c \int\frac{d^4 k}{(2\pi)^4} 
{\rm tr}_D\left[
\gamma_5 \gamma_\mu S_{f_1}(k+\hat\eta P) 
\Gamma_M(k;P) S_{f_2}(k - \eta P)\right]\,.
\end{eqnarray}
With this normalisation, $f_\pi \simeq 131\,$MeV.  Equation~(\ref{lep}) is
exact but approximation enters in the calculation of the Schwinger functions
that appear in it; i.e., the quark propagators and Bethe-Salpeter amplitude.

{\bf 2.1}~{\it Heavy-quark limit.} At this point, no assumption has been made
about the form of the Schwinger functions in Eqs.~(\ref{semilep}) and
(\ref{lep}).  In a study of light-meson decays, these functions were
determined from model DSEs, and the integrals then evaluated~\cite{KMR97}.
For heavy mesons, where the application of DSEs has not hitherto been
extensive, we will explore the consequences of Eqs.~(\ref{hqa}) and
(\ref{hqvtx}), in an estimation of the behaviour of the form factors and
decay constants: an exploration of a heavy-quark limit in our framework.

To be explicit, we set $\eta=1$ in Eqs.~(\ref{semilep}) and (\ref{lep})
(recall $\hat\eta=1-\eta$), so that the heavy-quark carries all the momentum
of the heavy meson, in which case the Bethe-Salpeter amplitude constrains the
momentum of the light-quark in the meson.  We write the heavy-meson momentum
as $P_\mu= (\hat M_{f_Q} + E)\, v_\mu$, where $E\equiv M_H - \hat M_{f_Q}$,
with $M_H$ the mass of the heavy meson, and $v_\mu$ is a timelike
unit-vector, $v^2= -1$.  Equation~(\ref{hqa}) then yields
\begin{equation}
\label{hqb}
S_{f_Q}(p-P) =  - \frac{1}{2}\,\frac{1 + i\gamma\cdot v}{p\cdot v + E} 
        + {\rm O}\left(\frac{|p|}{\hat M_{f_Q}},
                \frac{E}{\hat M_{f_Q}}
\right)\,.
\end{equation}
The bound state amplitudes, present in the integrands that arise in the
calculation of physical processes, ensure that $|p|/\hat M_{f_Q} \ll 1$.

Using Eq.~(\ref{hqa}), or (\ref{hqb}), the dressed quark-W-boson vertex is
given by Eq.~(\ref{hqvtx}).  For the heavy-meson Bethe-Salpeter
amplitude we employ the Ansatz 
\begin{equation}
\label{hqbsa}
\Gamma_{B^0,D^-}(k;P) = 
        \gamma_5 \left( 1 - \sfrac{1}{2} i\gamma\cdot v \right)
                \frac{ \varphi(k^2)}{ {\cal N}_{B^0,D^-}}\,,
\end{equation}
which is motivated by the Bethe-Salpeter equation studies of
Ref.~\cite{Bsep}.  This function describes the momentum distribution of the
light-quark, which we assume to be the same in both $B$ and $D$ mesons.  Our
estimate of the leptonic decay constants and form factors will be robust if
the results are insensitive to the form of $\varphi(k^2)$.

Using Eqs.~(\ref{hqvtx}), (\ref{hqb}) and (\ref{hqbsa}), the normalisation
condition, Eq.(\ref{bsanorm}), can be written
\begin{equation}
\label{bsanormH}
{\cal N}_H^2 = \frac{1}{M_H}\,\frac{N_c}{32 \pi^2}
\int_0^\infty du \,
\varphi(z)^2\,
\left(\sigma_S^f(z) + \sqrt{u} \,\sigma_V^f(z)\right)
\equiv \frac{1}{M_H \kappa_f^2} \,,
\end{equation}
where $z= u - 2 E \sqrt{u}$ and $H= B^0,D^-,\,$\ldots, etc., with the
light-quark flavour, $f$, chosen appropriately.  Clearly $ {\cal N}_{H_1^f}^2
M_{H_1^f} = {\rm constant} = {\cal N}_{H_2^f}^2 M_{H_2^f}$.  Similarly, we
find for a heavy-meson leptonic decay constant:
\begin{equation}
f_{H} = \frac{\kappa_f}{ \sqrt{M_{H}}}\,
        \frac{N_c}{8\pi^2}\,
        \int_0^\infty\,du\,
        (\sqrt{u}-E)\,\varphi(z)\,
        \left(\sigma_S^f(z) 
        + \sfrac{1}{2} \sqrt{u} \sigma_V^f(z)\right)\,,
\end{equation}
which entails that
\begin{equation}
\label{fhrule}
f_H \sqrt{M_H} = {\rm constant}\,,
\end{equation}
except for deviations due to differences between $u/d$- and $s$-quark
propagation characteristics.

Considering Eq.~(\ref{semilep}), we find, at leading order in $1/M_H$,
\begin{equation}
\label{fpm}
f_\pm (t) = \frac{1}{2} \,\frac{M_D \pm M_B}{\sqrt{M_D M_B}} \,\xi(w)\,,
\end{equation}
\begin{equation}
\label{xiwfn}
\xi(w) = \kappa_d^2 \frac{N_c}{32\pi^2}\int_0^1 d\tau\,\frac{1}{W}\,
\int_0^\infty du\, \varphi(z_W)^2\,
        \left(\sigma_S^d(z_W) + \sqrt{\frac{u}{W}}\, \sigma_V^d(z_W)\right)\,,
\end{equation}
with $W= 1 + 2 \tau (1-\tau) (w-1)$, $z_W= u - 2 E \sqrt{u/W}$ and
\footnote{The minimum physical value of $w$ is $w_{\rm min}=1$, which
corresponds to maximum momentum transfer with the final state meson at rest;
the maximum value is $w_{\rm max} \simeq (M_B^2 + M_D^2)/(2 M_B M_D) = 1.6$,
which corresponds to maximum recoil of the final state meson with the charged
lepton at rest.}
\begin{equation}
w = \frac{M_B^2 + M_D^2 - t}{2 M_B M_D} = v_B \cdot v_D\,.
\end{equation}
The canonical normalisation of the Bethe-Salpeter amplitude automatically
ensures that $\xi(w=1) = 1$.
Equation~(\ref{fpm}) is an example of a general result that, in the
``heavy-quark limit'' the semileptonic decays of heavy mesons are described
by a single, universal function~\cite{IW90}.

Equation~(\ref{xiwfn}) yields the result: $ \rho^2 \geq \rho^2_{\rm min}$,
where\footnote{For $\varphi(z)$ and $\sigma_{V/S}^f(z)$ non-negative,
non-increasing, convex-up functions of their argument, which includes
$\varphi=\,$constant and a free-particle propagator, $-\xi^\prime_E(w)$ takes
its minimum value at $E=0$.}
\begin{eqnarray}
\rho^2_{\rm min}
\equiv \left.- \frac{d\xi(w)}{dw}\right|_{w=1}^{E=0} & = &
\frac{1}{3} \left( 1 + 
\frac{\int_0^\infty\,du\,\varphi(u)^2 
        \,\sqrt{u}\,\sigma_V^d(u)}
        {2\int_0^\infty du\, \varphi(u)^2\,
        \left(\sigma_S^d(u) + \sqrt{u}\, \sigma_V^d(u)\right)}
        \right)\,,
\end{eqnarray}
which entails an upper bound on the slope of $\xi(w)$: $\rho^2 \gsim
\frac{1}{3}$, independent of the details of our model and consistent with
Ref.~\cite{BLRW97}.

{\bf 2.1}~{\it Light-quark propagator.}  The leptonic decay constant and
$\xi(w)$ depend on the light-quark propagator and the momentum distribution
of the light-quark in the heavy meson, described by $\varphi(k^2)$.  Many
observables involving hadrons composed of light-quarks have been studied
using the $u/d$- and $s$-quark propagators specified by
\begin{eqnarray}
\label{SSM}
\bar\sigma^f_S(x)  & =  & 
        2 \bar m_f {\cal F}(2 (x + \bar m_f^2))
        + {\cal F}(b_1 x) {\cal F}(b_3 x) 
        \left( b^f_0 + b^f_2 {\cal F}(\epsilon x)\right)\\
\label{SVM}
\bar\sigma^f_V(x) & = & \frac{2 (x+\bar m_f^2) -1 
                + e^{-2 (x+\bar m_f^2)}}{2 (x+\bar m_f^2)^2},
\end{eqnarray}
where ${\cal F}(y)\equiv (1-{\rm e}^{-y})/y$, $x=p^2/(2 D)$ and:
$\bar\sigma_V^f(x) = 2 D\,\sigma_V^f(p^2)$; $\bar\sigma_S^f(x) = \sqrt{2
D}\,\sigma_S^f(p^2)$; $\bar m_f$ = $m_f/\sqrt{2 D}$, with $D$ a mass scale.
This form is motivated by extensive studies of the DSE for the dressed-quark
propagator~\cite{DSErev} and combines the effects of confinement and
dynamical chiral symmetry breaking with free-particle behaviour at large
spacelike-$p^2$; i.e., asymptotic freedom.

The parameters $\bar m_f$, $b_{1\ldots 3}^f$ in Eqs.~(\protect\ref{SSM}),
(\protect\ref{SVM}) were determined in a $\chi^2$-fit to a range of
light-hadron observables, which is described in Ref.~\cite{BRT96} and leads
to the values in Table~\ref{tableA}.  In the fit, the difference between the
$u$- and $d$-quarks was neglected and only that minimal difference between
$u$- and $s$-quarks allowed that was necessary to ensure: $\langle \bar s
s\rangle <\langle \bar u u\rangle$; and $m_s/m_u\gg 1$.  ($\epsilon=10^{-4}$
is included in Eqs.~(\protect\ref{SSM}), (\protect\ref{SVM}) only for the
purpose of separating the small- and intermediate-$p^2$ behaviour of the
algebraic form, characterised by $b_0$ and $b_2$; a separation in magnitude
observed in numerical studies.)  These light-quark propagators have since
been used to successfully explore and predict a range of other hadronic
observables~\cite{KMR97,PL97}.  We use them herein.
\begin{table}[h,t]
\begin{center}
\caption{\label{tableA} The values of $b_{1,3}^s$ are underlined to indicate
that the constraints $b_{1,3}^s=b_{1,3}^u$ were imposed in the fitting.  The
scale parameter $D=0.160\,$GeV$^2$.}

\begin{tabular}{cccccc}\hline
    & $\bar m_f$ & $b_0^f$ & $b_1^f$ & $b_2^f$& $b_3^f$ 
\\\hline
$u$ & 0.00897 & 0.131 & 2.90 & 0.603 & 0.185 \\
$s$ & 0.224   & 0.105 & \underline{2.90} & 0.740 & \underline{0.185}
\\\hline
\end{tabular}
\end{center}
\end{table}

{\bf 3. Calculated heavy-meson observables}.\\ In this study we have
introduced one free parameter, $E\equiv M_H - \hat M_{f_Q}^E$, which is a
measure of the binding energy in a heavy meson, and an Ansatz for the
Bethe-Salpeter amplitude, Eq.~(\ref{hqbsa}), which involves a single
function, $\varphi(k^2)$.  Our calculated form of $\xi(w)$, and the
heavy-meson leptonic decay constants, will depend on these quantities.  Our
predictions will be robust if they are independent of the details of our
Ansatz for $\varphi(k^2)$.

\begin{figure}[t]
\centering{\
\epsfig{figure=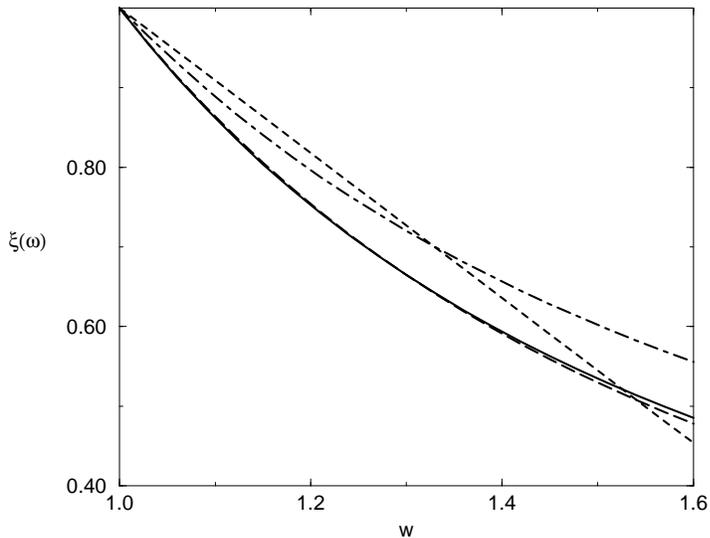,height=8.5cm}}
\caption{Our calculated form of $\xi(w)$ compared with experimental
fits~\protect\cite{CESR96}.  Experiment: dashed line,
Eq.~(\protect\ref{CESRfit}); short-dashed line, the linear fit $\xi(w) = 1 -
\rho^2 (w - 1)$, $\rho^2 = 0.91 \pm 0.15 \pm 0.06$.  Calculation: solid line,
$\xi(w)$ calculated with forms $A$, $B$ or $C$ in
Eq.~(\protect\ref{formsphi}) using the parameters in
Eq.~(\protect\ref{results}), there is no discernible difference on the scale
of this figure; dot-dash line, $\xi(w)$ calculated with form $D$.
\label{xiw}}
\end{figure}

To explore this we considered the following four one-parameter forms:
\begin{equation}
\label{formsphi}
\begin{array}{cccccccccc}
\varphi_A(k^2) && &\varphi_B(k^2) & & &\varphi_C(k^2) & &&
\varphi_D(k^2) \\\hline
\displaystyle \exp\left(-\frac{k^2}{\Lambda^2}\right)
& &&\displaystyle  \frac{\Lambda^2}{k^2+\Lambda^2}
& &&\displaystyle  \rule{0mm}{10mm}
\left(\frac{\Lambda^2}{k^2+\Lambda^2} \right)^2
& &&\displaystyle  \theta\left(1 - \frac{k^2}{\Lambda^2}\right)\,.
\end{array}
\end{equation}
For a particular form, to fit the parameters $(E,\Lambda)$, we performed a
$\chi^2$ fit to the following parametrisation of the experimental
data~\cite{CESR96}
\begin{equation}
\label{CESRfit}
\xi(w)=\frac{2}{w+1} \exp\left((1 - 2 \rho^2)\frac{w-1}{w+1}\right)\,,\;
        \rho^2 = 1.53\pm 0.36 \pm 0.14\,,
\end{equation}
to $f_D = 216 \pm 15\,$MeV and $f_B = 206 \pm 30\,$MeV, which, in the absence
of experimental data, is our weighted average of the lattice-QCD results
collected in Ref.~\cite{RB95}.  Using $M_D= 1.87\,$GeV, $M_{D_s}= 1.97\,$GeV
and $M_B=5.27\,$GeV, we obtain the results presented in Fig.~\ref{xiw} and
Eq.~(\ref{results}), energies in GeV and $\rho^2$ dimensionless.
\begin{equation}
\label{results}
\begin{array}{lc|cll|cllll}
  &&&  E   &  \Lambda  &&  f_D  &  f_{D_s} &  f_B     &  \rho^2\\\hline
A &&& 0.640 & 1.03  && 0.227  &  0.245  &  0.135  &  1.55 \\
B &&& 0.567 & 0.843 && 0.227  &  0.239  &  0.135  &  1.56 \\
C &&& 0.612 & 1.32  && 0.227  &  0.242  &  0.135  &  1.55 \\
D &&& 0.643 & 1.02  && 0.272  &  0.296  &  0.162  &  1.21 
\end{array}
\end{equation}

{\bf 4. Observations and Conclusions}.\\One observes from Eq.~(\ref{results})
that, independent of $\varphi(k^2)$, a description of the data requires a
``binding energy'' in $B$, $D$ mesons of $E \simeq 0.6\,$GeV, which is
consistent with the values obtained in Bethe-Salpeter equation studies of
these systems: $E \sim 0.4-0.6$~\cite{MJ92,Vary}.  Further, the
size-parameter $\Lambda \sim 1\,$GeV; and comparing this with an analogous
parametrisation of the pion Bethe-Salpeter amplitude shows $\Lambda \sim
(1.5\,$-$\,2.0)\, \Lambda_\pi$~\cite{ABR95}.  Hence, our Bethe-Salpeter
amplitude represents the heavy meson as on object of small spacetime extent;
occupying an intrinsic volume of only $5\,$-$\,20$\% that of the pion.

Globally, considering Eq.~(\ref{results}) and Fig.~\ref{xiw}, the Ans\"atze:
$A$,$B$,$C$, for the Bethe-Salpeter amplitude provide equivalent descriptions
of all observables.  The step-function, Ansatz $D$, which is qualitatively
different in form and corresponds simply to the introduction of a cut-off in
the integrals that yield observables, not surprisingly leads to a poorer
description.  The fact that even this form is not completely unreasonable is
due to the presence of the light-quark propagators in our study, which
describe the confinement of light-quarks in the heavy meson and restrict
their propagation range.  It is the {\em combination} of the light-quark
propagators, determined in studies of light-hadron observables, {\em and} the
heavy-meson Bethe-Salpeter amplitudes that prescribe the values of the
observables; i.e., as observed in Ref.~\cite{SS96}, these observables are
strongly influenced by the nonperturbative characteristics of confinement,
dynamical chiral symmetry breaking and bound state formation.

In our simple analysis, the calculated values of the leptonic decay constants
are in {\em quantitative} agreement with the results in Ref.~\cite{MJ92}.  In
particular, with the realistic Ans\"atze: $A$,$B$,$C$, we note that
$f_{D_s}/f_D \approx 1.07$, whereas Eq.~(\ref{fhrule}) suggests the result
$f_{D_s}/f_D = \sqrt{M_D/M_{D_s}} = 0.97$.  This difference, $\sim 10$\%, is
an indication of the influence and magnitude of light-quark effects in
heavy-meson observables.  Further, we note that while our value of $f_D =
227\,$MeV, agrees with that found in numerical simulations of lattice-QCD,
our value of $f_B = 135\,$MeV is $\sim 35$\% lower, and there appears no way
to reduce this discrepancy.  Our quantitative agreement with Ref.~\cite{MJ92}
suggests that the value inferred from lattice simulations may be erroneous.

We obtain significant curvature in $\xi(w)$ on the physically accessible
domain in $B\to D\ell \nu$ decays: $1<w\lsim 1.6$; i.e., as illustrated in
Fig.~\ref{xiw}, a comparison of our calculated result with the linear fit of
Ref.~\cite{CESR96} is unfavourable.  In the present study, no choice of our
two parameters yields a straight line, and trying to fit the short-dashed
line in Fig.~\ref{xiw} leads to a value of $\xi(w_{\rm max}) \sim$ 20\%
higher than that inferred from experiment.  We note that using any of the
three, realistic, one-parameter Ans\"atze for the Bethe-Salpeter amplitude,
our calculated result is in {\em exact} agreement with the experimental fit
to the data, Eq.~(\ref{CESRfit}).  The agreement between our calculation and
the other nonlinear fits described in Ref.~\cite{CESR96}: $\exp(-\rho^2
(w-1))$, $\rho^2= 1.27 \pm 0.29 \pm 0.12$; and $(2/(w+1))^{2\rho^2}$,
$\rho^2= 1.42 \pm 0.32 \pm 0.13$, is not quite as good, with the deviation
reaching $\sim 5$\% at $w_{\rm max}$.  From this we conclude that $\rho^2$ in
the range $\sim 1.2\,$-$\,1.6$ is admissible, with $\rho^2 \simeq 1.55$ most
likely.

Our calculation is not alone in indicating significant curvature in $\xi(w)$;
there are qualitative similarities with the results of
Refs.~\cite{SS96,IM96,ElHady95}, for example.  However, our calculation is
unique in being able to encompass a value of $\xi(w_{\rm max})\lsim 0.5$ with
physically reasonable and internally consistent values of our two parameters,
$E$ and $\Lambda$.  The detailed, constrained description of the light-quark
component of the heavy meson, via its propagator and its momentum
distribution function (Bethe-Salpeter amplitude), is responsible for this.

A quantitative step in extending this study is a thorough analysis of the DSE
for a heavy-quark, such as being explored in Ref.~\cite{CJB97}, which could
provide the basis for an estimation of corrections to the ``heavy-quark''
limit expressed by Eqs.~(\ref{hqa}) and (\ref{hqb}).  One could then employ
the calculated heavy-quark propagator in Bethe-Salpeter equation studies of
heavy-meson bound states, perhaps in a manner analogous to Ref.~\cite{BL97}.
In this way $E$ and $\Lambda$, which are fitting-parameters herein, would be
{\em correlated} and their values {\em fixed} by the mass scale in the gluon
propagator.  This would lead to predictions of leptonic decay constants and
semileptonic form factors that are less dependent on existing data.

{\bf Acknowledgments}.  The work of P.M. and C.D.R. was supported by the US
Department of Energy, Nuclear Physics Division, under contract number
W-31-109-ENG-38. The calculations described herein were carried out using the
resources of the National Energy Research Supercomputer Center.


\end{document}